\documentstyle[preprint,aps]{revtex}

\begin{document}  
\draft 
  
\title{ Destroying the superfluid state
of fermionic atoms by the magnetic field
in Feshbach resonance experiments.}  
\author{G.M.Genkin$^*$.}
\address{ Physics Department and
Center for Polymer Studies, Boston University ,
              Boston, MA 02215.}

\maketitle 
                 
\begin{abstract}

   We 
show that the superfluid state (SF) of fermionic atoms exist only for the
magnetic fields lesser than the critical magnetic field. This critical
magnetic field is determined by the equality of the Zeeman energy splitting
for atoms with different spin projections to the energy gap of fermionic
atoms without a magnetic field. For magnetic fields usually using in 
Feshbach resonance experiments a SF state is lost because these fields
are much more than the critical field. We show that the transition
temperature $ T_c $
of the SF state decreases as the magnetic field increases.

\end{abstract}  

  
   The 
trapping and cooling of gases with Fermi statistics has become one of the
central areas of research within the field of ultracold atomic
gases. Much
progress has been made in the achievement
of degenerate regimes of trapped atomic Fermi gases [ 1 - 5]. The major
goals of studies of these systems is to observe a transition to a
paired - fermion superfluid state. There has been considerable interest
in achieving superfluidity in an ultracold trapped Fermi gas in which a
Feshbach resonance is used to tune the interatomic attraction by
variation of a magnetic field. The interactions which drive the pairing
in
these gases can be controlled using a Feshbach resonance, in which a
molecular level is Zeeman tuned through zero binding energy using an
external
magnetic field. Via a Feshbach resonance it is possible to tune the
strength
and the sign of the effective interaction between particles. In result,
magnetic - field Feshbach resonances provide the means for controlling
the
strength of cold atom interactions, characterized $ s $ - wave
scattering
length $ a $, as well as whether they are
effectively repulsive ( $ a > 0 $) or attractive ( $ a < 0 $ ).
Therefore, the tunability of interactions in fermionic atoms provides a
unique
possibility to explore the Bose - Einstein condensate to
Bardeen - Cooper -Schriever ( BEC - BCS ) crossover [ 6 - 8 ], an
intriguing interplay between the superfluidity of bosons and Cooper
pairing of fermions. A Feshbach resonance offers the unique possibility
to
study the crossover between situations governed by Bose - Einstein and
Fermi - Dirac statistics. When the scattering length $ a $ is positive
the
atoms pair in a bound molecular state and these bosonic dimers can form
a
Bose - Einstein condensate; when $ a $ is negative, one expects the
well - known BCS model for superconductivity to be valid.

 In this paper, we show that the superfluid ( SF ) state of degenerate
fermionic gases analogous to superconductivity exist 
only for magnetic fields $ B < B_{cr} $. We show that this critical
magnetic
field which destroys the SF state
is determined by the equality of the Zeeman energy splitting
$ \Delta_{Zee} = 2\mu_{mag} B $
 for atoms
with
different spin ( spin - up and spin - down ) projections
to the energy gap $ \Delta_0 $ of fermionic atoms for $ T = 0 $ without
a magnetic field.
For
values of the magnetic field usually using in Feshbach resonance
experiments
for cold fermionic atoms a SF state is lost because these fields are
much more than the critical field. For example, for $ ^{6}Li $
atoms [ 9 ] the Zeeman energy splitting is about [ 10 ] $ 76 MHz $ where
as
$ \Delta_0 < 20 kHz ( 1\mu K) $. Note that in conventional
  superconductors
a superconductivity is lost for ferromagnetic phase [ 11]
if the Zeeman energy splitting which is determined by the magnetization
exceeds $ \sqrt {2} \Delta_0 $.

  We consider an uniform gas of Fermi atoms with two hyperfine ( spin -
  up  $ \uparrow $
and spin - down  $ \downarrow $ ) states in a magnetic field. Our
  starting model can be described by a Hamiltonian
$$
H = \sum_{\bf p}[( \eta_p + \mu_{mag} B ) a^{+}_{{\bf p} \uparrow}
a_{{\bf p} \uparrow}  + ( \eta_p - \mu_{mag} B) a^{+}_{{\bf p} \downarrow}
a_{{\bf p} \downarrow}] - \frac{g}{V} \sum_{{\bf p},{\bf q}} a^{+}_{{\bf
    p} \uparrow } a^{+}_{{ -\bf p} \downarrow} a_{{ -\bf q} \downarrow}
a_{{ \bf q} \uparrow },  \eqno (1)
   $$
with the coupling constant $ g = \frac{ 4 \pi h^2 |a|}{m} $, volume $ V $.
Here a coupling constant corresponds the attractive ( $ a < 0 $ )
pairing
interaction; $ a_{\bf p \sigma} ( a^{+}_{\bf p \sigma}) $ represent
the annihilation ( creation ) operators of a Fermi atom with the kinetic energy
$ \varepsilon_p = \frac{p^2}{2 m} $, and we use the notation
$ \eta_p = \varepsilon_p - \mu $, near the Fermi surface $ \eta_p = v_F
( p - p_F) $ where $ v_F = \frac{p_F}{m} $ and $ \mu $ is the chemical
potential. The Zeeman energy in an external magnetic field $ \bf B $ is
$ -\beta \bf{\sigma} \bf B $, and corresponding terms for spins $\uparrow $
and $ \downarrow $ are $ \pm \mu_{mag} B $ where $ \mu_{mag} $ is the
atomic magnetic moment. In the Hamiltonian (Eq.(1))
we introduce the standard canonical transformation to the Bogolyubov
 quasiparticles
  $$
a_{{\bf p} \uparrow} = u_p b_{{\bf p} \uparrow} +
 v_p b^{+}_{-{\bf p} \downarrow},
  $$
  $$
a_{{\bf p} \downarrow} = u_p b_{{\bf p} \downarrow} -
 v_p b^{+}_{-{\bf p} \uparrow},        \eqno (2)
  $$
where the
coefficients $ u_p $ and $ v_p $ are real, depend only on $ |p| $. They
are
chosen from the condition [12] that the energy $ E $ of the system has a
minimum for a given entropy. We have
  $$
E  = \sum_{\bf p} ( \eta_p + \mu_{mag} B) [u_p^2 n_{{\bf p} \uparrow}
  + v_p^2 ( 1 - n_{{\bf p} \downarrow})] +
  ( \eta_p - \mu_{mag} B) [u_p^2 n_{{\bf p} \downarrow}
  + v_p^2 ( 1 - n_{{\bf p} \uparrow})] -
 $$
 $$
 \frac{g}{V}[\sum_{\bf p} u_p v_p ( 1 - n_{{\bf p} \uparrow} -
 n_{{\bf p} \downarrow})]^2.           \eqno (3)
 $$
Varying this expression with respect to the parameter $ u_p $ and using
the
relation $ u_p^2 + v_p^2 = 1 $, which the transformation coefficients
must be satisfy, the condition for a minimum is
$\frac{\delta E}{\delta u_p} = 0 $. Using this condition for Eq.(3)
we have the standard [12]
equation for the uniform energy gap
$$
\frac{g}{2 V} \sum_{\bf p} \frac{( 1 - n_{{\bf p} \uparrow} -
 n_{{\bf p} \downarrow})}{ \sqrt{ \Delta^2 + \eta_p^2}} = 1,    \eqno (4)
$$
where the energy gap
  $$
\Delta =\frac{g}{V} \sum_{\bf p} u_p v_p ( 1 - n_{{\bf p} \uparrow} -
 n_{{\bf p} \downarrow})        \eqno (5a)
$$
and
$$
u_p^2 = \frac{1}{2} ( 1 + \frac{\eta_p}{\sqrt { \Delta^2 + \eta_p^2}}),
v_p^2 = \frac{1}{2} ( 1 - \frac{ \eta_p}{ \sqrt{ \Delta^2 +
 \eta_p^2}}),    \eqno(5b)
$$
here the quasiparticle occupation numbers $ n_{{\bf p} \alpha} =
 b_{{\bf p}\alpha}^{+} b_{{\bf p}\alpha} $.
The energy of the elementary excitations can be find [12] from the
change of the energy $ E $ of the system when the quasiparticle
occupation
numbers are changing, i. e. by varying $ E $ with respect to
$  n_{{\bf p} \uparrow} $
and $ n_{{\bf p} \downarrow} $. Therefore, start from the equation
$$
\delta E = \sum_{\bf p} [ \varepsilon_{\uparrow}(p) \delta n_{{\bf p} \uparrow}
+ \varepsilon_{\downarrow}(p) \delta n_{{\bf p} \downarrow}],    
$$
we have
$$
\varepsilon_{\uparrow}(p) = \frac{ \delta E}{\delta n_{{\bf p}
    \uparrow}} = \sqrt{\eta_p^2 + \Delta^2} + \mu_{mag} B,      \eqno (6a)
$$
$$
\varepsilon_{\downarrow}(p) = \frac{ \delta E}{\delta n_{{\bf p}
    \downarrow}} = \sqrt{\eta_p^2 + \Delta^2} - \mu_{mag} B,      \eqno (6b)
$$
 The quasiparticle occupation numbers satisfy Fermi - Dirac statistics.
 They
are defined by the Fermi distribution formula and the energy of the
 elementary
excitations
 are determined also [11] by the magnetic field.

  In the usual BCS gap equation ( Eq.(4))
using cutting off [12] the logarithmic integral
at same $ \eta =\bar{\epsilon} $ we
have the well - known result for the energy gap $ \Delta_0(g) $ for
$ T = 0 $ and
 $ B = 0 $
$$
ln \frac{\bar{\epsilon}}{\Delta_0(g)} = \frac{  2 \pi^2 h^3}
{ g m p_F}.     \eqno (7a)
$$
 For $ B \neq 0 $, in general, the energy
gap
depends on the magnetic field $ \Delta(g,B)$.
The energy of the elementary excitations must be positive, therefore, if
$ \Delta < \mu_{mag} B $ there is a nonzero 
minimal value in the integral ( Eq.(4))
$ \eta_{p (min)} = \sqrt{(\mu_{mag} B)^2 - \Delta^2} > 0 $.
Using the table integral $ \int\frac{dx}{\sqrt{x^2 + a^2}} = ln | x +
 \sqrt{ x^2 + a^2} | $
for $ \Delta < \mu_{mag} B $
we have in this case
$$
ln \frac{\bar{\epsilon}}{ \mu_{mag} B + \sqrt{ (\mu_{mag} B)^2 - 
\Delta^2(g,B)}} = \frac { 2 \pi^2 h^3}{g m p_F}.     \eqno (7b)
$$
By comparing Eqs.(7) the energy gap $ \Delta(g,B) $ is determined by the
equation
$$
 \mu_{mag} B + \sqrt{ (\mu_{mag} B)^2 - 
\Delta^2(g,B)} = \Delta_0(g),      \eqno (8a)
$$
or
$$
\Delta(g,B) = \sqrt{\Delta_0(g) ( 2\mu_{mag} B - \Delta_0(g) )}.
\eqno (8b)
$$
This spectrum was obtained in Ref.[11] for the ferromagnetic phase of
the
conventional superconductors. The equation (8a) has a threshold at the
value
$ 2 \mu_{mag} B_{th} = \Delta_0(g) $ for which this equation first has
solutions  for $ B \succeq B_{th} $, and, accordingly, the energy gap
$ \Delta(g,B) $ has a solution only for   $ B \succeq B_{th} $.
At the threshold the Zeeman splitting $ \Delta_{Zee}^{th} =
 2 \mu_{mag} B_{th} $ is equal to the energy gap $ \Delta_0(g) $,
i. e. $ \Delta_{Zee} = \Delta_0 $.
 For
$ B < B _{th} $ the minimal value of $\eta_p $ in the BCS gap equation
equal to zero, i.e. $\eta_{p (min)} = 0 $, 
therefore, for these magnetic fields we have the standard gap parameter
$ \Delta_0(g)$. Beyond the threshold $ B > B_{th} $ there is 
$ \eta_{p(min)} \neq 0 $  and, therefore, the gap parameter is
$\Delta(g,B) $. At the threshold itself $ \Delta(g,B_{th}) = 0 $ and
$ \eta_{p(min)}(B_{th}) = \frac{\Delta_0}{2 \mu_{mag}} $.

 The BCS superfluid state of a two component Fermi gas is encoded in a
 single
BCS gap $ \Delta $. As a result, the energy gap for magnetic fields
$ \Delta_{Zee} < \Delta_{Zee}^{th} $ is $ \Delta_0(g) $, and
 for
magnetic fields above the threshold $ \Delta_{Zee} \succeq
 \Delta_{Zee}^{th} $
 the energy gap is $ \Delta(g,B) $.

 We consider the thermodynamic properties of Fermi atoms with two
 hyperfine
states for magnetic fields
 above the threshold. In calculations it is convenient
to start from the thermodynamic potential $ \Omega $, and the difference
between the thermodynamic potential $ \Omega_s $ in the superfluid state
 and
the value in the normal state $ \Omega_n $ at the same temperature [12]
 is
$$
 \Omega_s - \Omega_n = -\int_{0}^{g} \frac{\Delta^2(g_1)}{g_1} dg_1.
 \eqno (9)
$$
Changing in Eq.(9)
from integration over $ dg_1 $ to that over
$ d\Delta $, and using Eq.(8b) we obtain the difference between the
ground - state energies of the superfluid $ E_s $ and normal $ E_n $
states for magnetic fields $ 2 \mu_{mag} B \succeq \Delta_0 $, or
$ \Delta_{Zee} \succeq \Delta_0 $
 $$
E_s - E_n = \frac{ m p_F}{ 4 \pi^2 h^3} ( \Delta_{Zee} - \Delta_0 )^2,
\eqno (10a)
$$
for $ \Delta_0 \preceq \Delta_{Zee} <  2 \Delta_0 $; and
$$
E_s - E_n = \frac{ m p_F}{ 4 \pi^2 h^3}  \Delta_0^2,   \eqno (10b)
$$
for $ \Delta_{Zee} > 2 \Delta_0 $.
The positive sign of this difference $ E_s - E_n > 0 $
indicates that for
magnetic fields $ \Delta_{Zee} > \Delta_0 $ the normal state has the
lesser
energy. For magnetic fields $ \Delta_{Zee} < \Delta_0 $ this difference
is
negative, and we have a BCS state. Therefore, our system passes from
the BCS state to the normal state as the field increases to a critical
value
$$
B_{cr} = B_{th} = \frac{ \Delta_0}{ 2 \mu_{mag}},   \eqno (11a)
$$
i. e. a critical value of the Zeeman splitting $ \Delta_{Zee}^{cr} $ is
$$
 \Delta_{Zee}^{cr}   = \Delta_{Zee}^{th} = \Delta_0.    \eqno (11b)
$$
By means of Eq.(11) we have
$$
 \frac{ B}{B_{cr}} = \frac{\Delta_{Zee}}{\Delta_0}.     \eqno (12)
$$

 In  Feshbach
resonance
experiments for fermionic atoms $ \Delta_{Zee} \sim 80 MHz, \Delta_0 \prec 20
kHz $ that is why a SF state can not exist for
these
magnetic fields because of $ B \gg B_{cr} $, or $ \Delta_{Zee} \gg
\Delta_0 $.

   For $ B < B_{cr} $ there is a superfluid BCS state. The critical
temperature $ T_c $ is determined by standard manner [12], and in the
BCS gap equation ( Eq.(4)) the quasiparticle occupation numbers are
given by the Fermi distribution formula ( Eqs.(6)). Following [12],
we obtain the temperature dependence of the energy gap $ \Delta $
near the transition point $ T \preceq T_c $ for small magnetic field
$ \frac{ \mu_{mag} B}{ k T_c} \ll 1 $
  $$
ln \frac{ \Delta_0}{ \Delta} = ( 1 +
 \alpha ) [ ln \frac{ \pi T }{\gamma \Delta} + \frac{ 7 \xi(3)}{ 8
     \pi^2} \frac{\Delta^2}{ T^2} ], \alpha = 2 (\frac{\mu_{mag} B}
 { k T_c})^2 = \frac{1}{2} (\frac{ \Delta_{Zee}}{ k T_c})^2.   \eqno (13)
  $$
Hence, we see that the energy gap $ \Delta \rightarrow 0 $ 
at a temperature $ T_c $
  $$
( k T_c )^{1 + \alpha} = ( \frac{\gamma}{\pi})^{ 1 +
 \alpha} \Delta_0 \Delta^{\alpha}.     \eqno (14)
  $$
Note that our calculation (Eq.(14))
is valid for a small energy gap $\Delta \rightarrow 0 $, but a nonzero
gap $ \Delta \neq 0 $. For $ B = 0 $
we have the well - known expression for the critical temperature
$ k T_c = \frac{\gamma}{\pi} \Delta_0, \frac{\gamma}{\pi} = 0.57 $;
however, for $ B \neq 0 $
 the 
critical temperature $ T_c $ decreases.

  In summary,
we show that the superfluid state of fermionic atoms is destroying by the
magnetic field. Due to the Zeeman energy splitting for atoms with
different
spin projections the superfluid state exist only for the magnetic fields
lesser than the critical magnetic field.
As the magnetic field increases the transition temperature $ T_c $ of
the
superfluid state decreases.
 The critical magnetic field is
determined by the equality of the Zeeman energy splitting to the energy
gap without a magnetic field. For the magnetic field greater than the
critical field the energy gap depends on the magnetic field, however, for
these
magnetic fields the normal state has the lesser energy than the
superfluid state. For values of magnetic fields usually using in 
Feshbach resonance experiments a SF state is lost because these fields
are much more than the critical field.

   I thank H. E. Stanley for encouragement, J. Borreguero for assistance.


\begin{thebibliography}{20}
           \bibitem[*]{*} Electronic address: ggenkin@argento.bu.edu.  
           \bibitem{1} B. Demacro and D. S. Jin, Science {\bf 285},
 1703 (1999). 
         \bibitem{2} A. G. Truscott, K. E. Strecker, W. I. Mc Alexander,
       G. Partridge, and R. G. Hulet, Science {\bf 291}, 2570 (2001).
        
         \bibitem{3} F. Schreck, L. Khaykovich, K. L. Corwin,
         G. Ferrari,
T. Bourdel, J. Cubizolles, and C. Salomon,
  Phys. Rev. Lett. {\bf 87}, 080403 (2001).
         \bibitem{4} S. R. Granade, M. E. Gehm, K. M. O'  Hara, and
J. E. Thomas, Phys. Rev. Lett. {\bf 88}, 120405 (2002).
         \bibitem{5} Z. Hadzibabic, S. Gupta, C. A. Stan, C. H. Schunck,
M. W. Zwierlein, K. Dieckmann, and W. Ketterle, 
 Phys. Rev. Lett.
         {\bf 91}, 160401 (2003).
         \bibitem{6} E. Timmermans, K. Furuga, P.M. Milonni, and 
A. K. Kerman,
          Phys.  Lett. A{\bf 285}, 228 (2001).
         \bibitem{7} M. Holland, S. M. F. Kokkelmans, M. L. Chiofalo,  
 and
        R. Walser, Phys. Rev. Lett. {\bf 87}, 120406 (2001).
         \bibitem{8} Y. Ohashi and A. Griffin,
         Phys. Rev. Lett. {\bf 89}, 130402 (2002).
         \bibitem{9} T. Bourdel, L. Khaykovich, J. Cubizolles, J. Zhang, 
         F. Chevy, M. Teichman, L. Tarruell, S. M. F. Kokkelmans,                  and
        C. Salomon, Phys. Rev. Lett. {\bf 93}, 050401 (2004).
         \bibitem{10} Note that for bosonic atoms $ ^{85}Rb $ 
the Zeeman energy splitting is about $ 536 MHz $
( the JILA experiments, E. A. Donley {\it  et. al.}, Nature ( London ) 
{\bf 417}, 529 (2002), 
the dependence of the Zeeman splitting on the magnetic field is
$ h^{-1} \frac{\delta \Delta_{Zee}}{\delta B} = 34.6 \frac{MHz}{mT} $
and the position of the Feshbach resonance is $ B_0 = 15.49 mT $). 

         \bibitem{11} A. A. Abrikosov, {\it Fundamentals of the theory
         of metals} ( North - Holland, New York, 1988), Chapter 21.
         \bibitem{12} See, for example, E. M. Lifshitz and L. P. Pitaevskii,
  {\it Statistical Physics}, Part 2 ( Pergamon Press, New York, 1980 ).  1964 ),
        
                 
         
         
         
         
         
         
            
           
           \end{thebibliography}
\end{document}